\newcommand{\kT}{k_{\rm B}T}

\documentclass[10pt]{article}

\usepackage[dvips]{graphicx}
\usepackage{epsfig}
\usepackage{amstext}
\usepackage{amssymb}
\usepackage{amsmath}
\usepackage{amsmath}
\usepackage{amssymb}
\usepackage{amsfonts}
\usepackage{rotate}
\usepackage{rotating}
\usepackage{cite}
\usepackage{color} 
\usepackage{multirow}
\usepackage{enumerate}

\usepackage[labelfont=bf,labelsep=period,justification=raggedright]{caption}
\makeatletter
\renewcommand{\@biblabel}[1]{\quad#1.}
\makeatother

\date{}

\begin{document}

\begin{flushleft}
{\Large
\textbf{Exploring the Free Energy Landscape: From Dynamics to Networks and Back}
}
\\
Diego Prada-Gracia$^{1,2}$, 
Jes\'us G\'omez-Garde\~nes$^{2,3}$, 
Pablo Echenique$^{2,4}$,
Fernando Falo$^{1,2,\ast}$
\\
\bf{1} Departamento de F\'isica de la Materia Condensada, Universidad de Zaragoza, E-50009 Zaragoza, Spain
\\
\bf{2} Instituto de Biocomputaci\'on y F\'\i sica de Sistemas Complejos (BIFI), Universidad de Zaragoza, E-50009 Zaragoza, Spain
\\
\bf{3} Departamento de Matem\'atica Aplicada, ESCET, Universidad Rey Juan Carlos, E-28933 M\'ostoles (Madrid), Spain
\\
\bf{4} Departamento de F\'isica Te\'orica, Universidad de Zaragoza, E-50009 Zaragoza, Spain
\\
$\ast$ E-mail: fff@unizar.es
\end{flushleft}

\section*{Abstract}

The knowledge of the Free Energy Landscape topology is the essential
key to understand many biochemical processes. The determination of the
conformers of a protein and their basins of attraction takes a central
role for studying molecular isomerization reactions. In this work, we
present a novel framework to unveil the features of a Free Energy
Landscape answering questions such as how many meta-stable conformers
are, how the hierarchical relationship among them is, or what the
structure and kinetics of the transition paths are. Exploring the
landscape by molecular dynamics simulations, the microscopic data of
the trajectory are encoded into a Conformational Markov Network.  The
structure of this graph reveals the regions of the conformational
space corresponding to the basins of attraction. In addition, handling
the Conformational Markov Network, relevant kinetic magnitudes as
dwell times and rate constants, or hierarchical relationships among
basins, complete the global picture of the landscape. We show the
power of the analysis studying a toy model of a funnel-like potential
and computing efficiently the conformers of a short peptide, the
dialanine, paving the way to a systematic study of the Free Energy
Landscape in large peptides.

\section*{Author Summary}

A complete description of complex polymers, such as proteins, includes
information about their structure and their dynamics. In particular it
is of utmost importance to answer the following questions: What are
the structural conformations possible?  Is there any relevant
hierarchy among these conformers?  What are the transition paths
between them?  These and other questions can be addressed by analyzing
in an efficient way the Free Energy Landscape of the system. With this
knowledge, several problems about biomolecular reactions (such as
enzymatic activity, protein folding, protein deposition diseases, etc)
can be tackled. In this article we show how to efficiently describe
the Free Energy Landscape for small and large peptides. By mapping the
trajectories of molecular dynamics simulations into a graph (the
Conformational Markov Network), and unveiling its structural
organization, we obtain a coarse grained description of the protein
dynamics across the Free Energy Landscape in terms of the relevant
kinetic magnitudes of the system. Therefore, we show the way to bridge
the gap between the microscopic dynamics to the macroscopic kinetics by
means of a mesoscopic description of the associated Conformational
Markov Network. Along this path the compromise between the physical
nature of the process and the magnitudes that characterize the network
is carefully kept to assure the reliability of the results shown.

\section*{Introduction}

Polymers and, more specifically, proteins, show complex behavior at
the cellular system level, {\em e.g.} in protein-protein interaction
networks \cite{Oltvai2002Systems}, and also at the individual level, where
proteins show a large degree of multistability: a single protein can
fold in different conformational states \cite{Parisi1994Complexity,Krivov2002Free,Krivov2004Hidden}. As a
complex system \cite{Ottino2003Complex,Amaral2004Complex}, the dynamics of a
protein cannot be understood by studying its parts in isolation,
instead, the system must be analyzed as a whole. Tools able to
represent and handle the information of the entire picture of a
complex system are thus necessary.

Complex network theory \cite{Newman2003Structure,Boccaletti2006Complex} has proved to be a
powerful tool used in seemingly different biologically-related fields
such as the study of metabolic reactions, ecological and food webs,
genetic regulatory systems and the study of protein dynamics
\cite{Newman2003Structure}. In this latter context, diverse studies have
analyzed the conformational space of polymers and proteins making use
of network representations \cite{Noe2007Hierarchical, Rao2004Protein, Caflisch2006Network,Noe2008Transition}, where nodes
account of polymer conformations. Additionally, some studies have
tried to determine the common and general properties of these
conformational networks \cite{Scala2001Smallworld,Gfeller2007Uncovering} looking at magnitudes
such as clustering coefficient, cyclomatic number, connectivity,
etc. Recently, trying to decompose the network in modules
corresponding to the free energy basins, the use of community
algorithms over these conformational networks have been proposed
\cite{Gfeller2007From}. Although this approach has opened a promising path for
the analysis of Free Energy Landscapes (FEL), the community based
description of the network leads to multiple characterizations of the
FEL and thus it is difficult to establish a clear map from the
communities found to the basins of the FEL.

A similar approach, commonly used to analyze the complex dynamics, is
the construction of Markovian models. Markovian state models let us
treat the information of one or several trajectories of molecular
dynamics (MD) as a set of conformations with certain transition
probabilities among them
\cite{Chodera2007Automatic,Noe2007Hierarchical,Buchete2008Coarse}. Therefore,
the time-continuous trajectory turns into a transition matrix,
offering global observables as relaxation times and modes. In
\cite{Chodera2007Automatic,Deuflhard2000Identification,Buchete2008Coarse}
the use of Markovian models is proposed with the aim of detecting FEL
meta-stable states. However, the above approaches to analyze FELs of
peptides involves extremely large computational cost: either general
community algorithms or large transition matrices.

Finally, other strategies to characterize the FEL that have
successfully helped to understand the physics of biopolymers, are based
on the study of the Potential Energy Surface (PES)
\cite{Krivov2002Free,Krivov2004Hidden,Wales2006Energy,Evans2003Free,Wales2000Energy}. The classical
transition-state theory \cite{Hanggi1990Reactionrate} allows us to project the
behavior of the system at certain temperature from the knowledge of
the minima and transition states of the PES. This approach entails
some feasible approximations, such as harmonic approximation to the
PES, limit of high damping, assumption of high barriers, etc. These
approximations could be avoided working directly from the MD data.

In this article we make a novel study of the FEL capturing its
mesoscopic structure and hence characterizing conformational states
and the transitions between them. Inspired by the approaches presented in
\cite{Gfeller2007From,Noe2008Transition} and \cite{Chodera2007Automatic,Buchete2008Coarse}, we translate a
dynamical trajectory obtained by MD simulations into a Conformational
Markov Network. We show how to efficiently handle the graph to obtain,
through its topology, the main features of the landscape: conformers
and their basins of attraction, dwell times, rate constants between
the conformational states detected and a coarse-grained picture of the
FEL. The framework is shown and validated analyzing a synthetic
funnel-like potential. After this, the terminally blocked alanine
peptide (Ace-Ala-Nme) is studied unveiling the main characteristics of
its FEL.

\section*{Methods}

In this section we show the round way of the FEL analysis: the map of
microscopic data of a MD into a Conformational Markov Network (CMN)
and, by unveiling its mesoscopic structure, the description of the FEL
structure in terms of macroscopic observables.

\subsection*{Translating the FEL into a network}

First, we encode a trajectory of a stochastic MD simulation into a network:
the CMN. This map will allow us to use the tools introduced henceforth
to analyze a specific dynamics of complex systems such as biopolymers.

\subsubsection*{Conformational Markov Network}

The CMN has been proven to be a useful representation of large
stochastic trajectories \cite{Rao2004Protein,Caflisch2006Network,Gfeller2007From}. This coarse
grained picture is usually constructed by discretizing the
conformational space explored by the dynamical system and considering
the hops between the different configurations as dictated by the MD
simulation. In this way, the nodes of a CMN are the subsets of
configurations defined by the conformational space discretization and
the links between nodes account for the observed transitions between
them. The information of the stochastic trajectory allows to assign
probabilities for the occupation of a node and for the transitions
between two different configurations. Defined as above, a CMN is thus
a weighted and directed graph.

Our CMN is constructed as follows. The conformational space is divided
into $N$ cells of equal volume, therefore every node $i$ ($i=1,...,N$)
of the CMN contains the same number of possible configurations. Next,
by evolving a stochastic trajectory enough time steps (of length
${\Delta t}$) to assure the ergodicity condition we can define the
final CMN set up. We assign to each node a weight, $P_{i}$, that
accounts for the fraction of trajectory that the system has visited
any of the configurations contained in node $i$ (the following
normalization $\sum_{i} P_{i}=1$ holds). Second, a value $P_{ij}$ is
assigned to each directional link, accounting for the number of hops
from node $j$ to node $i$. Note that transitions between
configurations contained in the same node are also considered by
$P_{ii}$, {\em i.e.}  the network can also contain
self-loops. Finally, the weights of the outgoing links from a node
$j$, $\{P_{ij}\}$, are conveniently normalized so that $\sum_{i}
P_{ij}=1$.

The CMN constructed in this way, is described by a single matrix
$\mathbf{S}=\{P_{ij}\}$ and a vector whose components are the
occupation probabilities $\vec{P}=\{P_{i}\}$. Hence, the matrix
$\mathbf{S}$ is the transition probability matrix of the following
Markov chain,
\begin{equation}\label{evolves}
\vec{v}(t+\Delta t)=\mathbf{S} \vec{v}(t)
\end{equation}
where $\vec{v}(t)$ is the instant probability distribution of the
system at time $t$. Since the matrix $\mathbf{S}$ is ergodic and time
invariant, one can compute the stationary distribution associated to
the Markov chain, $\vec{v}^{e}$, that satisfies
$\vec{v}^{e}=\mathbf{S} \vec{v}^{e}$. The latter stationary
distribution has to be identical to the computed weights of the
network nodes, $P_{i}=v^{e}_{i}$ ($i=1,...,N$), provided the
stochastic trajectory is long enough. Moreover,
 the detailed balance condition, 
\begin{equation}\label{balance}
P_{ji}P_{i}=P_{ij}P_{j}
\end{equation}
must hold thus relating the elements of matrix $\mathbf{S}$ to the
stationary probability distribution. Therefore, the transition matrix
$\mathbf{S}$ appears to be the minimal descriptor of the stochastic
trajectory and, as consequence, of the CMN.

\subsubsection*{Markovity}

Provided the MD trajectory is long enough to consider the sample in
equilibrium, the weight-distribution of nodes in the CMN will be the
stationary solution of Eq. (\ref{evolves}) and detailed balance
condition (\ref{balance}) will be fulfilled
\cite{VanKampen2001Stochastic}. However, this property is not enough
to consider the model Markovian: Although the continuous trajectory
will be produced using Langevin dynamics (and therefore inherently
Markovian in the phase space
\cite{VanKampen1998Remarks,Zwanzig2001Nonequilibrium}) the discrete
representation of the CMN and the integration of momenta defies the
Markovian character of our model
\cite{VanKampen1998Remarks,Jernigan2003Testing,Swope2004Describing,Zwanzig1983From}. Several
methods are proposed in the literature to validate Markov models
\cite{Park2006Validation,Chodera2007Automatic,Swope2004Describing}. In
order to obtain a reliable description, specially about those
magnitudes related to the time evolution of the system (see subsection
Temporal hierarchy of basins), the time step ${\Delta t}$ must be
large enough to avoid memory effects \cite{Swope2004Describing}.

A detailed check and discussion about the Markovian character of the
networks shown in this article can be found in the \emph{Supporting Information Text (SI)}.

\begin{center}
\begin{figure}[!b]
\begin{center}
\includegraphics[width=0.9\textwidth]{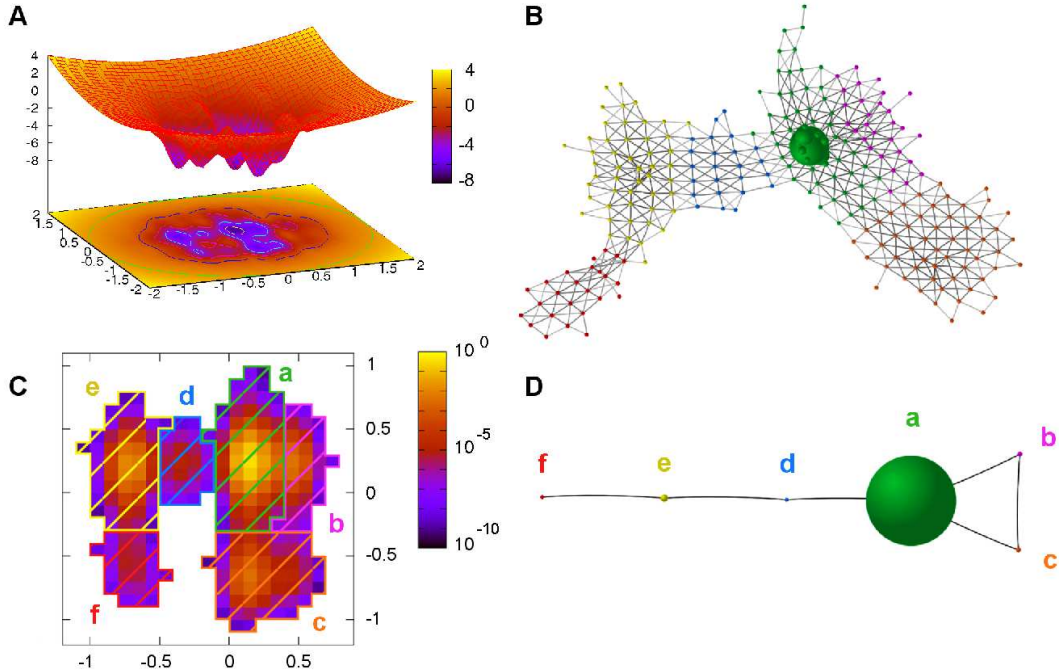}
\end{center}
\caption{ {\bf SSD algorithm applied to a synthetic funnel-like
    potential.} (A) 2D funnel-like potential. (B) A stochastic
  trajectory is translated into a CMN where 6 sets of nodes
  (corresponding to different color) are the result of the SSD
  algorithm. (C) Recovering the spatial coordinates, the stationary
  probabilities of each node are shown in color code. The 6 basins
  detected are represented as color striped regions. (D) A
  coarse-grained CMN is built where new nodes take the role of the
  basins.}\label{funnel}
\end{figure}
\end{center}

\subsubsection*{Funnel-like potential}\label{section_funnel}

To illustrate the CMN approach and the methods presented below, we
introduce here a synthetic potential energy function, that serve us as
a toy model where results can be easily interpreted. This potential
energy is reminiscent of that funnel surfaces recurrently found when
the FEL of proteins are studied
\cite{Wolynes1995Navigating,Onuchic2004Theory}. In particular, we have
considered a two-dimensional system where a particle moves in contact
with a thermal reservoir and whose potential energy is given by,
\begin{eqnarray}\label{eq:funnel}
V(x,y) & = & \frac{1}{2} (x^{2}+y^{2})-(a_{0}+a_{1}\sin(b_{1}x)+a_{2}\sin(b_{2}x)  
\nonumber 
\\
& & +a_{3}\sin(b_{3}y))e^{-\frac{1}{2}(x^{2}+y^{2})^{2}}\;,
\end{eqnarray}
where we have set $a_{0}=4$, $a_{1}=1.6$, $a_{2}=0.8$, $a_{3}=1.2$,
$b_{1}=6$, $b_{2}=15$ and $b_{3}=6$. As shown in Figure \ref{funnel}A
the above potential energy confines the movement of the stochastic
trajectory inside a finite region of the conformational
space. However, thermal fluctuations allow the particle to jump
between several basins of attraction.

A stochastic trajectory has been simulated using an overdamped
Langevin dynamics and the equations of motion have been integrated with a fourth
order stochastic Runge-Kutta method \cite{Greenside1981Numerical}. Figure
\ref{funnel}C shows the region of the conformational space visited by
the particle. We have conveniently discretized the two-dimensional space into
pixels of equal area and computed their corresponding occupation
probabilities. Thereby, with the transition probabilities between
pixels, the trajectory is represented as the CMN shown in
Figure \ref{funnel}B. The question now is: can we recover the topology
of the FEL (derived from Eq. (\ref{eq:funnel})) from the CMN
representation?

\subsection*{Analyzing the FEL through the network}

Up to now, we have illustrated the conversion of molecular dynamics
data into a graph (the CMN). Now, we show how to efficiently obtain
the thermo-statistical data from the mesoscopic description of the CMN.

\subsubsection*{Revealing structure: conformational basins}

Inspired by the deterministic steepest descent algorithm to locate
minima in a potential energy surface we propose a \emph{Stochastic
  Steepest Descent} (SSD) algorithm to define basins on the
discretized FEL. Dealing with the nodes and links as we describe
below, the proper structure of the CMN is unveiled to call the modules
obtained conformational macro-states or basins.

Picking at random one node of the CMN, say $a$, and an initial
probability distribution $P_{i}({0})=\delta_{i,a}$ ($i=1,...,N$), the
Markov process relaxes according to $\vec{P}({\Delta
  t})=\mathbf{S}\vec{P}({0})$. The whole probability concentrated in
node $a$ at time $0$, in a single time step $\Delta t$, evolves
driving the maximum amount of probability down hill over the FEL. The
next node $b$ in the descendent pathway from $a$ is taken by following
the link that carries maximum probability flux. Focusing again all the
probability in node $b$, $P_{i}(1)=\delta_{i,b}$ we continue the
pathway from $a$ towards a local FEL minimum by identifying the next node
$c$ for which the probability current $P_{c,b}$ is maximal. Iterating
this operation for each node of the CMN, we obtain a set of
disconnected descent pathways that help us to define the basins of
attraction.

We establish formally the above procedure assisted by a vector
$\vec{\Omega}=\{\omega_{i}\}$ (with $i=1,...,N$) that label the nodes:

\begin{itemize}

\item[{\em (i)}] We start by assigning $\vec{\Omega}=\vec{0}$.

\item[{\em (ii)}] Select at random a node $l$ with $\omega_{l}=0$
  ({\em i.e.}, $l$ has not been labeled yet) and start to write an
  auxiliary list $V$ of nodes adding $l$ as the first entry in this
  list.
 
\item[{\em (iii)}] Search, within the neighbors of the node $l$, a
  node $m$ so that $P_{ml}=max\{P_{j,l},\forall j\neq l
  \}$\cite{nota1} and check which of the following options is
  fulfilled:

\begin{enumerate}[A]

\item{If $P_{ml}>P_{lm}$ and $\omega_{m}=0$: add $m$ to the list $V$ and go
  again to {\em (iii)} taking $m$ in the place of $l$.}

\item{If $P_{ml}>P_{lm}$ and $\omega_{m}\neq 0$ then write the
  labels of all the nodes in the list $V$ as
  $\omega_{j}=\omega_{m}$ $\forall j\in V$. The process continues going to
  step {\em (ii)}.}

\item{If $P_{ml}\leq P_{lm}$ the link $l\to m$ is removed from the
    graph. The process returns to step {\em (iii)} with the next
    exception: since this step has been iterated $2D$ times for the
    same node $l$ (being $D$ the number of coordinates discretized to
    construct the CMN), $l$ is stated as local minimum and
    $\omega_{l}=l$. In this case $\omega_{j}=l$ for those nodes $j \in
    V$ and the process comes back to step {\em (ii)}.}
\end{enumerate}

\end{itemize}

The whole procedure ends when no nodes unlabeled remain in the CMN,
$\omega_i\neq 0$ $\forall i$. The restriction introduced in step ({\em
  iii}.C) with the dimensionality $D$ avoids a transition from a local
minimal energy configuration to any other node of the same basin or to
a deeper local minimum of a different basin.  When every node of the
CMN has been visited, the conformational space is completely
characterized and we have thus traced all the maximum descent pathways
from any node to the local FEL minima. Finally, all those nodes with
the same label $\omega_{i}$ belong to the same FEL basin and therefore
they are associated to the same conformational state. The result of
the procedure is the partition of the CMN in a set of modules which
correspond to basins of attraction of the discretized conformational
space.

To illustrate the basin decomposition of a CMN, the SSD algorithm has
been applied to the funnel-like potential. The result is the detection
of six basins in agreement with the number of local minima in its FEL
(Figure \ref{funnel}B and \ref{funnel}C).

\subsubsection*{Comparing with other community algorithms}

With the aim of studying biomolecules and systems with high degree of
dimensionality, the way to detect these FEL basins must be
computationally efficient. The method described above takes a
computational time $\mathcal{O}(2DN)$, once the $2D$ largest hooping
probabilities $P_{ji}$ are computed for all the nodes in the
network. Additionally, the method is deterministic providing with a
unique partition of the CMN into different modules. These two
characteristics make this analysis faster and more straightforward
than any other partitioning method \cite{Danon2005Comparing}. These advantages
come from the knowledge of the physical meaning of links and nodes of
CMNs.  In the \emph{SI} other community algorithms (Newman's
modularity and Markov Clustering algorithm) are tested over our toy
model system. None of the algorithms reported in the \emph{SI} give a
satisfactory result mapping the modules obtained with the free energy
basins.

\subsubsection*{Coarse-Grained CMN}
 
To get a more comprehensible representation of the FEL studied, a new
CMN network can be built by taking the basins as nodes. The occupation
probabilities of these nodes as well as the transition probabilities
among them can be obtained from those of the original CMN as
\begin{equation}
P_{\alpha}=\sum_{i \in \alpha} P_{i}\;,
\end{equation}
\begin{equation}
P_{\beta \alpha}=\frac{\sum_{i \in \alpha} \sum_{j \in \beta} P_{ji} P_{i}}{\sum_{i \in \alpha} P_{i}} \;,
\end{equation}

where $i$ and $j$ are indexes relative to the nodes of the original
CMN and $\alpha$ and $\beta$ are indexes for the basins (new
nodes). Note that the new coarse-grained CMN has its weights
normalized and fulfills the detailed balance condition
Eq. (\ref{balance}). Figure \ref{funnel}D shows the corresponding
coarse-grained for the funnel-like potential.

The weighted nodes and links have a clear physical meaning
\cite{Wales2006Energy}. Considering the transition state $\alpha \rightarrow
\beta$ and assuming local "intra-basin" equilibrium, the rate constant
of this transition is $k_{\alpha \beta}=P_{\beta \alpha}/ \Delta t$
(where $\Delta t$ is the time interval between snapshots used to make
the original CMN). The relative free energy of the basin $\alpha$,
taking basin $\beta$ as reference, is $\Delta F_{\alpha}=-\kT
log(P_{\alpha}/P_{\beta})$. Besides, the expected waiting time to
escape from $\alpha$ to any adjacent basin is $\tau_{\alpha}={\Delta
  t}/(1-P_{\alpha \alpha})$ \cite{Trygubenko2006Graph}. Other magnitudes, such
as first-passage time for inter-basins transitions and other rate
constants relaxing the local equilibrium condition
\cite{Wales2006Energy,Trygubenko2006Graph} can also be computed from the original CMN.

The ability to define the proper regions of the conformational space
in an efficient way let us compute physical magnitudes of
relevance. For instance, the coarse-grained CMN is nothing but a
graphical representation of a kinetic model with $n$ (the number of
basins) coupled differential equations:
\begin{equation}\label{eq:diff}
\frac{dP_{\alpha}(t)}{dt}= -\sum_{\beta \neq \alpha} k_{\beta \alpha} P_{\alpha}(t) + \sum_{\beta \neq \alpha} k_{\alpha \beta} P_{\beta}(t).
\end{equation}

\begin{center}
\begin{figure}[!b]
\begin{center}
\includegraphics[width=0.60\textwidth]{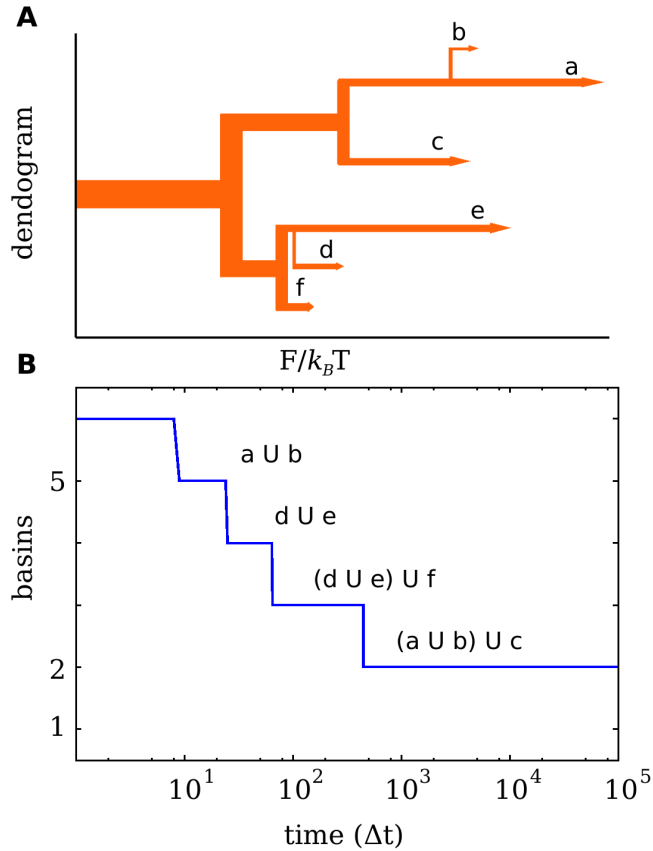}
\end{center}
\caption{{\bf Hierarchies of the basins detected for the funnel-like
      potential.} (a) Free Energy hierarchy: based on the relative free-energy
    of the nodes. (b) Temporal hierarchy: number of basins defined by SSD
    for the different networks built by eq. (\ref{eq:time}). The
    original basins merge in function of time. Both hierarchies
    reveals a coarse-grained behavior of two macro-states: $(a \cup b
    \cup c)$ and $(d \cup e \cup f)$.}
\label{dendo}
\end{figure}
\end{center}

\subsubsection*{Free Energy hierarchical basin organization}

The first hierarchy aims to answer the following question: What is the
structure of the CMN when nodes with lower weight than a certain
threshold are removed together with their links? Let us take the
control parameter $F/\kT$ as the threshold to restrict the existence
of the nodes in the original CMN. Where
$F_{i}/\kT=log(P_{w})-log(P_{i})$ is the \emph {``adimensional free
energy''} of a node $i$ relative to the most weighted node $w$.

With the above definitions we start a CMN reconstruction by smoothly
increasing the threshold from its zero value. At each step of this
process, we obtain a network composed of those nodes with free energy
lower than the current threshold value. As the free-energy cut-off
increases, new nodes emerge together with their links. These new nodes
may be attached to any of the nodes already present in the network or
they can emerge as a disconnected component. At a certain value of
$F/\kT$, some components of the network become connected
by the links of a new node incorporated at this step. A
Hoshen-Kopelman like algorithm \cite{Hoshen1976Percolation} is used to
detect the disconnected components of the network at each value of the
threshold used: from zero until all the $N$ nodes of the CMN were
already attached.

This bottom-up network reconstruction provides us with a hierarchical
emergence of nodes along with the way they join together. This picture
can be better described by a process of basins emergence and linking
that is easily represented by means of a basin dendogram. This
representation let us guess at first glance the hierarchical
relationship of the conformational macro-states and the height of the
barriers between them. Let us remark that the transition times cannot
be deduced from these qualitative barriers since the entropic
contribution or the volume of the basin are not reflected in this
diagram. The basins family-tree obtained for the funnel-like (see
Figure \ref{dendo}A) reveals that, despite of having a two-dimensional
potential with the shape of a funnel, one cannot describe it as a
sequence of metastable conformations that drive the system to the
global minimum. Moreover, the diagram shows a roughly similar behavior
as for a double asymmetric well, composed by two sets of basins: $(a
\cup b \cup c)$ and $(d \cup e \cup f)$.

\subsubsection*{Temporal hierarchy of basins}\label{temporal_hierarchy}

The CMN representation of a MD simulation provides with another
hierarchical relationship that is meaningful to understand the
behavior of the biological systems. The links of the original CMN
have been weighted according to the stochastic matrix
$\mathbf{S}=\{P_{ij}\}$. Taking into account the Markovian character
of the process, we can make use of the Chapman-Kolmogorov equation to
generate new transition matrices at times $\tau=2\Delta t$, $3\Delta
t$, etc... Formally, the Markov chain at sampling time $m\Delta t$ is
defined by the matrix:

\begin{equation}\label{eq:time}
\mathbf{S}(m\Delta t)=[\mathbf{S}(\Delta t)]^{m}\;.
\end{equation}

For each value of $m$ a new CMN is defined. This family of CMNs have
different weighted links but the same weights $P_{i}$ for the nodes as
the original one ($m=1$). It is worth to discuss the behavior of the
matrix $\mathbf{S}(m\Delta t)$. In the limit $m\rightarrow \infty$ we
have $\vec{P}=\mathbf{S}(m\Delta t) \vec{P}(0)\;,$ independently of
the initial state $\vec{P}(0)$. This means that any node is connected
to a given node of the network with the same weight, regardless of the
initial source. Therefore, only one basin would be detected by the SSD
algorithm since every node is connected with the most weighted link to
the most weighted node in only one step.

From the original $\Delta t$-description of the FEL to the integrated
($ m\Delta t \rightarrow \infty$) one, we can devise another algorithm
to establish a second hierarchy of the basins by performing the next
two operations: First, for each value of $m$ a new CMN is generated by
constructing the matrix $[S(\Delta t)]^{m}$ and second, the SSD
algorithm is applied to this new CMN. The process finishes when only a
basin (the whole network) is detected (for large enough values of
$m$). By using this technique we can observe how basins merge with
others at different time scales (labeled by the integer $m$).

The result of this procedure performed for the funnel-like potential
is shown in Figure \ref{dendo}B. At $\tau \approx 500$ only two basins
are observed: $(a \cup b \cup c)$ and $(d \cup e \cup f)$, being the
largest plateau observed for any of the nontrivial arrangement of
basins found. Therefore, the macroscopic description in time is in
agreement with the Free Energy hierarchy described previously. It is
clear that the number of basins decrease as $m$ increases. One should
be aware that the concept of basin depends dramatically on the time
resolution at which the CMN is built, and this time limits also the
resolution in the FEL structure. Note also that this procedure
provides with useful information similar to the \emph{structural
  decorrelation time} \cite{Lyman2007Structural}.

\begin{center}
\begin{figure}[!b]
\begin{center}
\includegraphics[width=0.60\textwidth]{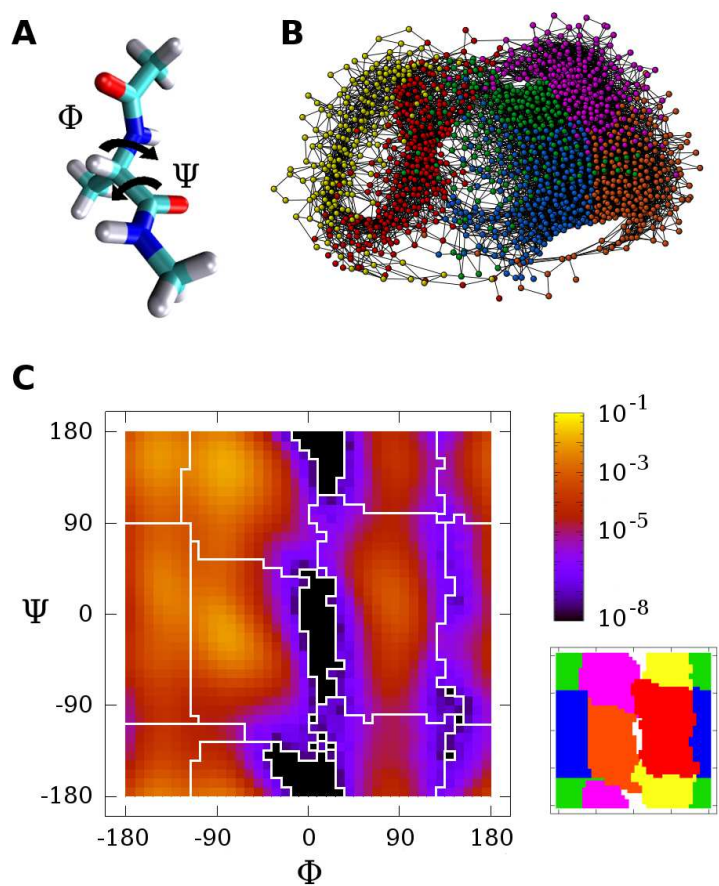}
\end{center}
\caption{{\bf Free energy basins of the Alanine
    dipeptide.}(A) The dialanine dipeptide with the angles $\phi$ and
  $\psi$. (B) Plot of the CMN generated. The $6$ sets of nodes
  (corresponding to different colors) are the result of the SSD
  algorithm. (C) Left: Ramachandran plot with the probability of
  occupation of the cells used to build the CMN. The boundaries of the
  free energy basins are shown in white. Right: the $6$ basins
  represented as regions of different color. (Color code: orange =
  $\alpha_{R}$, red = $\alpha_{L}$, yellow = $C_{7ax}$, pink =
  $C_{7eq}$, green = $C_{5}$ and blue = $\alpha'$).}
\label{dipeptido}
\end{figure}
\end{center}

\section*{Results}
\subsection*{The Alanine dipeptide.}\label{alanina}

The alanine dipeptide, or terminally blocked alanine peptide
(Ace-Ala-Nme, Figure \ref{dipeptido}A), is the most simple "biological molecule'' that
exhibits the common features shown by larger biomolecules. Despite of
its simplicity, this system has more than one long-life conformational
state with different transition pathways.
Since the first attempt by Rossky and Karplus \cite{rossky79} to model
this dipeptide solvated, this system has been widely studied in
theoretical works \cite{Smith1999Alanine,Apostolakis1999Calculation,Chekmarev2004LongTime,Bolhuis2000Reaction}. The
alanine dipeptide has been also the appropriate molecule to test tools
to explore the FEL \cite{Hummer2003Coarse,Chodera2007Automatic,Gfeller2007From} and, specifically,
to study reaction coordinates \cite{Bolhuis2000Reaction,Ma2005Automatic}.

The alanine dipeptide has two slow degrees of freedom, the rotatable
bonds $\phi$ (C-N-C$_{\alpha}$-C) and $\psi$ (N-C$_{\alpha}$-C-N) (see
Figure \ref{dipeptido}A). The FEL projected onto these dihedral angles
let us identify the conformational states that characterize the
geometry of biopolymers, namely: alpha helix right-handed
($\alpha_{R}$), alpha helix left-handed ($\alpha_{L}$), beta strands
($C_{7eq}$,$C_{5}$), etc. The number of local minima in the ($\phi$,
$\psi$) space depends on the effective potential model used to
simulate the system. Up to date, electronic structure calculations
have identified a total of nine different conformers
\cite{Vargas2002Conformational}. Regarding MD simulations different conformational
states have been observed: {\em (i)} using classical force fields with
explicit solvent up to six conformers are detected
\cite{Chodera2007Automatic,Apostolakis1999Calculation,Smith1999Alanine}, {\em (ii)} at least four stable
states by using implicit solvent \cite{Gfeller2007From,Smith1999Alanine,Chekmarev2004LongTime}, and
{\em (iii)} two stable conformers in vacuum conditions
\cite{Bolhuis2000Reaction,Smith1999Alanine}. On the other hand, since the angles $\phi$ and
$\psi$ seem appropriate to distinguish the metastable states, the
kinetics between them is not accurately described with this choice of
reaction coordinates, the solvent coordinates and/or other internal
degrees of freedom must be taken into account \cite{Ma2005Automatic,Bolhuis2000Reaction}.

We have used \emph{SSD} algorithm to detect the local minima and their
corresponding basins for this molecule in the $\phi$-$\psi$ space. For
this purpose, a Langevin MD simulation of $250$ ns has been performed
at a temperature of $400$ K (see the \emph{SI} for further
details). Additionally the CMN has been built dividing the $(\phi
,\psi)$ Ramachandran plot into cells of surface $9^{ \circ}\times 9^{
  \circ}$ ($40\times 40$) and taking dialanine conformations at time
intervals of $\Delta t=0.01$ ps. The resulting CMN have a total of
$n=1505$ nodes and $e=26324$ directed links.

The $SSD$ algorithm applied to the CMN network reveals $6$
basins. Figure \ref{dipeptido}B shows the resulting network where nodes
belonging to the same basin take the same color.
Bringing back this information to the Ramachandran map, these $6$ sets
of nodes define $6$ regions represented in Figure \ref{dipeptido}C. To
better illustrate this division, other representation,
where each region has a different color, is shown. By comparing
with previous studies on this molecule, we identify the regions in orange, red,
yellow and pink with conformers $\alpha_{R}$, $\alpha_{L}$,
$C_{7ax}$ and $C_{7eq}$ respectively \cite{Smith1999Alanine,Roterman1989Comparison}. Besides,
region green corresponds to conformer $C_{5}$ and the blue one to
$\alpha'$ \cite{Chodera2007Automatic,Roterman1989Comparison}. Remarkably, the basin
$\alpha_{L}$ (one of the less populated state) has been visited $1155$
times with a mean stay time of $4.20$ ps.

We now look at the coarse-grained picture of the FEL by describing the
properties of the $6$ basins detected. The different weights of the
basins are related to the free energy of the corresponding
conformational macro-states. In Table \ref{MST} these energy
differences $\Delta F_{a}= - \kT log(P_{a}/P_{C_{7eq}})$ are shown,
taking the most populated basin as the energy reference
$F_{C_{7eq}}=0$. The lowest free energy basins correspond to
configurations with $\phi \leq 0^{ \circ}$ ($C_{7eq}$, $\alpha_{R}$, $C_{5}$
and $\alpha'$), whereas the two other conformers located in the region
$\phi \geq 0^{ \circ}$ have the highest free energy but the largest dwell time.
Moreover, we have also analyzed the trapping efficiency of each basin
by reporting the mean escape time ($\Delta t/(1-P_{aa})$) as well in Table
\ref{MST}.

\begin{center}
\begin{table}[!htpb]
  \caption{{\bf Relative free energies and Mean Escape Time of the basins defined by SSD.}}
\begin{center}
\begin{tabular}{l c c}
\hline
Basin & $F_{i}-F_{C_{7eq}}$ kcal/mol & Mean Escape Time (ps) \\
\hline
\hline
$C_{7eq}$ & 0.00 & 0.52 \\
$\alpha_{R}$ & 0.45 & 0.42 \\
$\alpha_{L}$ & 2.42 & 4.20 \\
$C_{7ax}$ & 3.84 & 0.71 \\
$C_{5}$ & 0.55 & 0.28 \\
$\alpha'$ & 0.90 & 0.23 \\
\hline
\end{tabular}
\end{center}
\label{MST}
\end{table}

\begin{table}[!htpb]
  \caption{ {\bf Characteristic times for direct inter-basins transitions.}}
\begin{center}
\begin{tabular}{l c | l c}
\hline
$a\to b$ & $1/k_{ba}$ (ps) & & \\
\hline
\hline
$C_{7eq}\to \alpha_{L}$ & 1968.34 & $\alpha_{L}\to C_{7eq}$ & 88.24 \\
$\alpha_{R}\to C_{7ax}$ & 58011.74 & $C_{7ax}\to \alpha_{R}$ & 815.87 \\
\hline
$C_{5}\to C_{7ax}$ & 393.75 & $C_{7ax}\to C_{5}$ & 6.63 \\
$\alpha'\to \alpha_{L}$ & 400.57 & $\alpha_{L}\to \alpha'$ & 58.47 \\
\hline
$C_{7eq}\to \alpha_{R}$ & 3.32 & $\alpha_{R}\to C_{7eq}$ & 1.88 \\
$\alpha_{L}\to C_{7ax}$ & 4.80 & $C_{7ax}\to \alpha_{L}$ & 0.78 \\
\hline
\end{tabular}
\label{transitions}
\end{center}
\end{table}

\end{center}

\begin{center}
\begin{figure}[!hb]
\begin{center}
\includegraphics[width=0.60\textwidth]{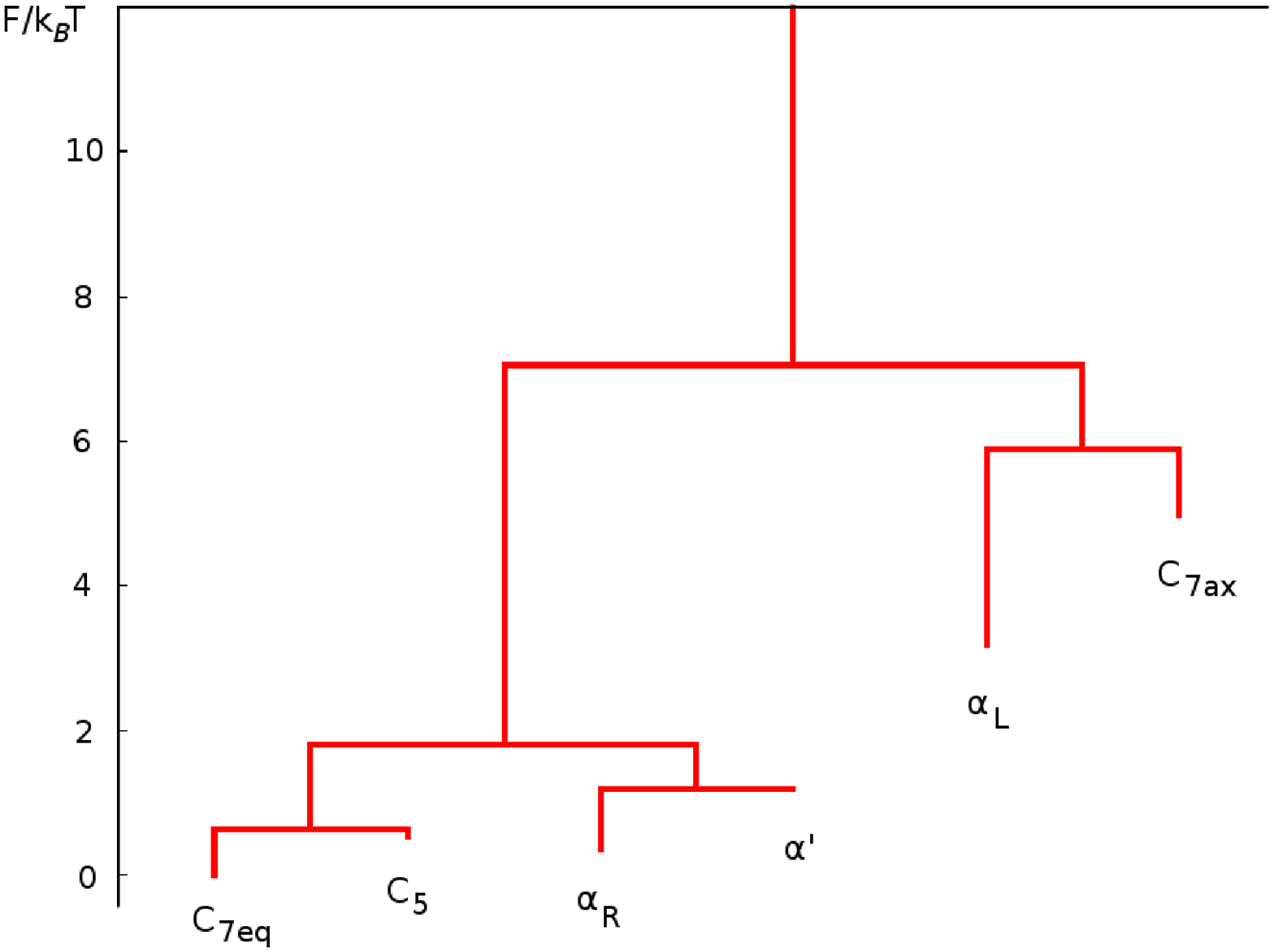}
\end{center}
\caption{{\bf Dendogram based on the relative Free Energy of the CMN
    nodes.} Two sets of basins are clearly distinguished with a high
  free energy barrier in between: ($C_{7eq}$, $\alpha_{R}$, $C_{5}$,
  $\alpha'$) and ($C_{7ax}$,$\alpha_{L}$). Note that $\alpha_{L}$
  looks like the conformer with the largest dwell time, in agreement
  with data in Table \ref{MST}. }
\label{dia_dendo}

\begin{center}
\includegraphics[width=0.60\textwidth]{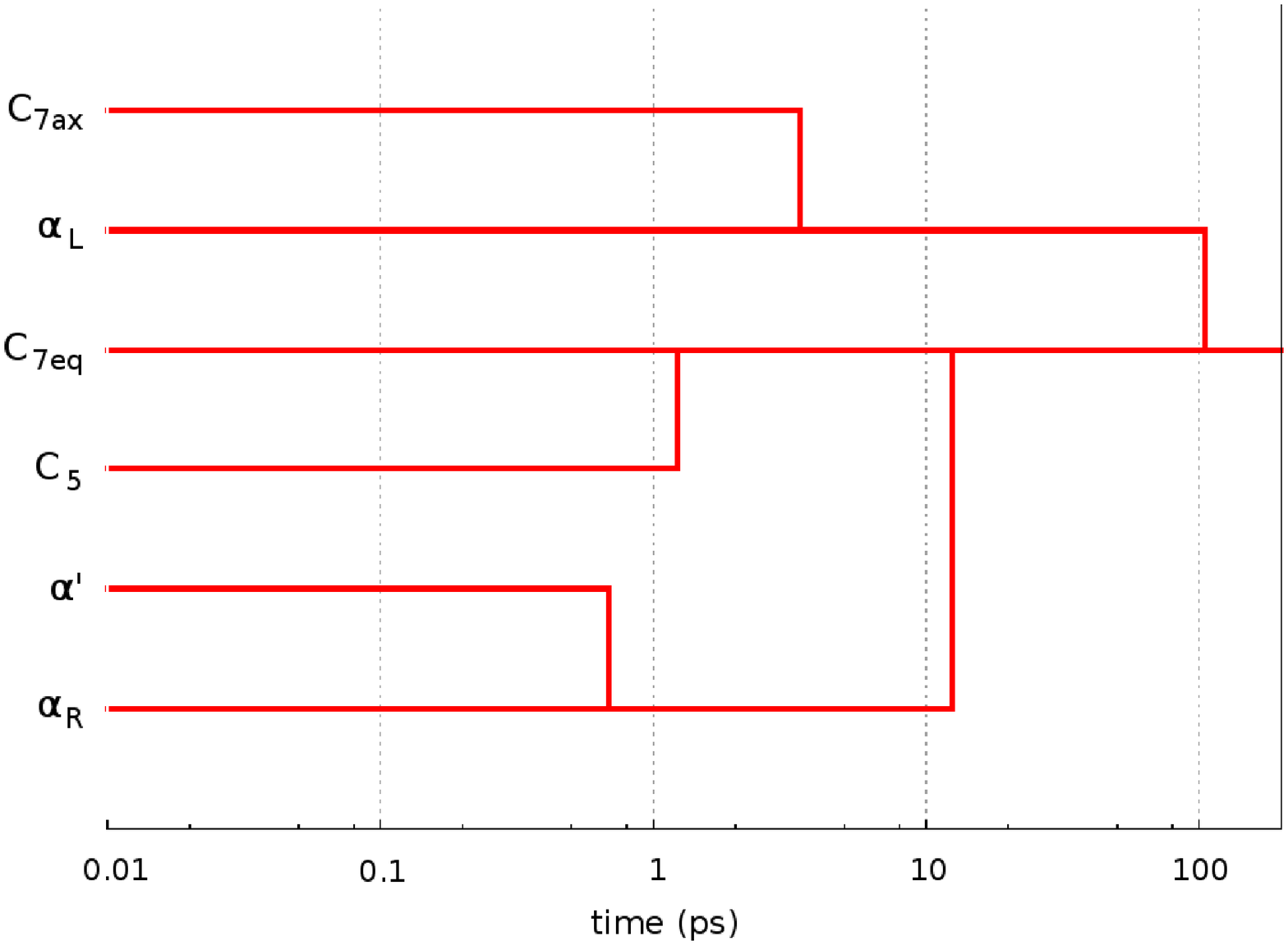}
\end{center}
\caption{{\bf Dendogram based on the temporal hierarchy of basins.} In
  around 100 ps the peptide finds the way to reach the global minimum,
  conformer $C_{7eq}$, from any basin.}
\label{dia_dendo2}
\end{figure}
\end{center}

The FEL can be represented as a dendogram, see Figure \ref{dia_dendo},
where the hierarchical map of the conformers based on Free Energy
gives at first glance a global picture of the landscape. Remarkably,
the conformer $\alpha_{L}$, despite of having one of the highest free
energy, looks like the metastable state with longest life. This result
is supported by the values of Mean Escape Time shown in Table
\ref{MST}.

The alanine dipeptide has been also studied because of its "fast''
isomerization $C_{7eq}\to \alpha_{R}$ and a "slow'' transition
$\alpha_{R}\to C_{7ax}$. Our coarse-grained picture of the FEL also
allows us to extract information about these transitions. In Table
\ref{transitions} we show some of the relevant characteristic
transition times from a basin \emph{a} to an adjacent basin \emph{b},
$1 /k_{ba}$. [The whole data is shown in the \emph{SI}.]  Transitions
between basins with the same sign of $\phi$ are remarkably faster
({\em e.g} $C_{7eq}\leftrightarrow \alpha_{R}$ and
$\alpha_{L}\leftrightarrow C_{7ax}$). While slow transitions are
observed for those hops crossing the line $\phi=0^{ \circ}$ ($C_{7eq}\to
\alpha_{L}$ and $\alpha_{R}\to C_{7ax}$), showing them as rare
events. Instead, the alanine dipeptide finds more easily paths to go
to $\phi \geq 0^{ \circ}$ conformers through $C_{5}\to C_{7ax}$ and $\alpha'\to
\alpha_{L}$ by crossing $\phi=180^{\circ}$.

To round off the description of the FEL, the dendogram corresponding
to the temporal hierarchy is shown in Figure \ref{dia_dendo2}. From the
figure, it becomes clear that the behavior of the dialanine depends on
the time scale used for its observation. Again, the same two different
sets of conformers are distinguished from this hierarchy. Additionally,
the global minimum conformer is reached in around 100 ps from any basin.

Finally, the magnitudes computed here for the alanine dipeptide would
allow to construct a first-order kinetic model of $6$ coupled differential
equations as Eq. (\ref{eq:diff}) (assuming equilibrium
intra-basin). This model contains the same information as the kinetic
model by Chekmarev et al. for the irreversible transfer of population
from $\alpha_{R} \to C_{7ax}$ \cite{Chekmarev2004LongTime}.

\section*{Discussion}

Hierarchical landscapes characterize the dynamical behavior of
proteins, which in turn depends on the relation between the topology
of the basins, their transitions paths and the kinetics over energy
barriers.  The CMN analysis of trajectories generated by MD
simulations is a powerful tool to explore complex FELs. 

In this article, we have proposed how to deal with a CMN to unveil the
structure of the FEL in a straightforward way and with a remarkable
efficiency. The analysis presented here is based on the physical
concept of basin of attraction, making possible the study of the
conformational structure of peptides and the complete characterization
of its kinetics. Note that this has been done without the estimation of the
volume of each conformational macro-state in the coordinates space and
without the 'a priori' knowledge of the saddle points or the
transition paths from a local minimum to another.

On the other hand, the framework introduced in the article provides us
with a quantitative description of the dialanine's FEL, coming up
directly from a MD dynamics at certain temperature.  The peptide
explores its landscape building the corresponding CMN and the success
of extracting the relevant information is up to the ability of dealing
with it. Neither the FE basins were defined by the unique criterion of
clustering conformations with a geometrical distance \cite{Becker1997Geometric},
nor the rate constants were projected from the potential energy
surface \cite{Wales2006Energy,Evans2003Free}. Moreover, the conformers and their
properties were computed from the MD with the only limitation of the
discretization of time and space.

Although we have applied the method to low dimensional landscapes, we
expect that high dimensional systems could be also studied, by the
combination of this technique with the usual methods to reduce the
effective degrees of freedom (like principal component analysis or
essential dynamics). In conclusion, the large amount of information
obtained by working with the CMN, its potential application to any
peptide with a large number of monomers, and the possibility of
performing the analysis on top of CMN constructed via several short MD
simulations \cite{Chodera2006LongTime}, make the approach presented here a
promising way to describe the FEL of a protein.

\section*{Acknowledgments}
A critical reading of the manuscript by Y. Moreno and L.M. Flor\'ia,
and the helpful comments and suggestions from the anonymous referees
are gratefully acknowledged.

\newpage

\title{\bf Supporting Information }

\maketitle

\section{Checking Markovity}

\begin{figure*}[hb!]
\begin{center}
\includegraphics[width=0.50\textwidth,angle=270]{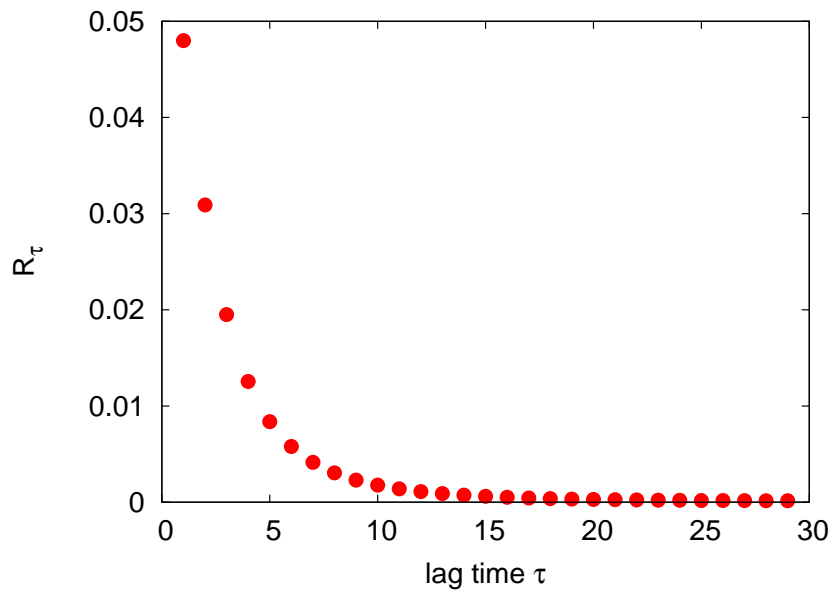}
\end{center}
\caption{\label{FigureS1} Checking Markovity: The relative mutual information
  $R_{\tau}$ quantifies the degree of
  non-Markovity of our stochastic model. The lag time used to
  construct the CMN -funnel like potential-, $\tau=1$, reveals a
  memory effect lower than a $5\%$. }
\end{figure*}

Although the purpose of the work introduced in the manuscript was to
illustrate a way to handle the Conformational Markov Network (CMN) in
order to obtain the basins of attraction and characterize the Free
Energy Landscape, the lector must be advise about the necessity of
checking the markovian character of the model (network in our case).
\\
\\
With regard to the CMNs that appear in the manuscript:
\\
\\
(i)\hspace{1cm} The particle in the {\em funnel like potential} is
simulated using an overdamped dynamics. In this case, the continuous
trajectory integrated is inherently Markovian \cite{VanKampen1998Remarks2,
  Zwanzig2001Nonequilibrium2}. When we discretize the coordinate space to lump the
trajectory into micro-states, the Markovity of the new description can
be defied \cite{Huisinga2003Extracting2,Jernigan2003Testing2}. Therefore, one has to care about
the time step to describe the process since it must be larger than the
longest equilibration time among the micro-states \cite{Swope2004Describing2}. In
ref. \cite{Chodera2007Automatic2} three approaches are mentioned to evaluate the
degree of Markovity of a stochastic model. Among these approaches, we
have taken the criterion presented by Park and Pande in \cite{Park2006Validation2}
based on Shannon's entropy. This method provides a unique magnitude
not as sensitive to statistical and numerical noise as other
methods. Another reason to make this choice is that a necessary and
sufficient condition for Markovity is checked (while the observation
of the eigenvalues of the transition matrix is not a sufficient
condition).  The analysis is based on the comparison of a first-order
conditional entropy $H_{\tau}(X_{n}|X_{n-1})$ and the second-order one
$H_{\tau}(X_{n}|X_{n-1},X_{n-2})$ (see \cite{Park2006Validation2} for further details
and notation). The magnitude $R_{\tau}$ quantifies, given $X_{n-1}$,
what fraction of the information in $X_{n}$ is the mutual information
between $X_{n}$ and $X_{n-2}$. Although the method is computationally
expensive, our models are small enough to carry out this analysis. The
Figure \ref{FigureS1} reveals that with the lag time used in our
analysis ($\tau =1$) the memory effect (non-Markovity) is less than a
$5\%$.

We have to remark, as it is discussed in ref. \cite{Park2006Validation2}, that
there's probably no answer to the question: when does the model
behaves Markovian? However the appropriate question should be: Given
an observable, what is the degree of Markovity needed to have a
correct measure? Regarding to our work, note that the analysis of the
basins detection depends on the detailed balance condition, but not on
the relaxation times. On the other hand, a small deviation such as the
$5\%$, only affects at the analysis of the temporal hierarchy of
basins, where Chapman-Kolmogorov is used explicitly. In order to
distinguish whether the error is propagated (and increased) or not
when the transition matrix is raised to the power of $n$, we compute
the Kullback-Leibler divergence between the transition matrix
$S(\tau)^{n}$ and the matrix computed from the trajectory $S(n \tau)$:

\begin{displaymath}
  D_{KL}(S(n \tau)||S(\tau)^{n})=\sum_{ij}S(n \tau)_{ij}P_{j} \ln \frac{S(n \tau)_{ij}}{[S(\tau)^{n}]_{ij}}
\end{displaymath}

\begin{figure*}[hb!]
\begin{center}
  \includegraphics[width=0.50\textwidth,angle=270]{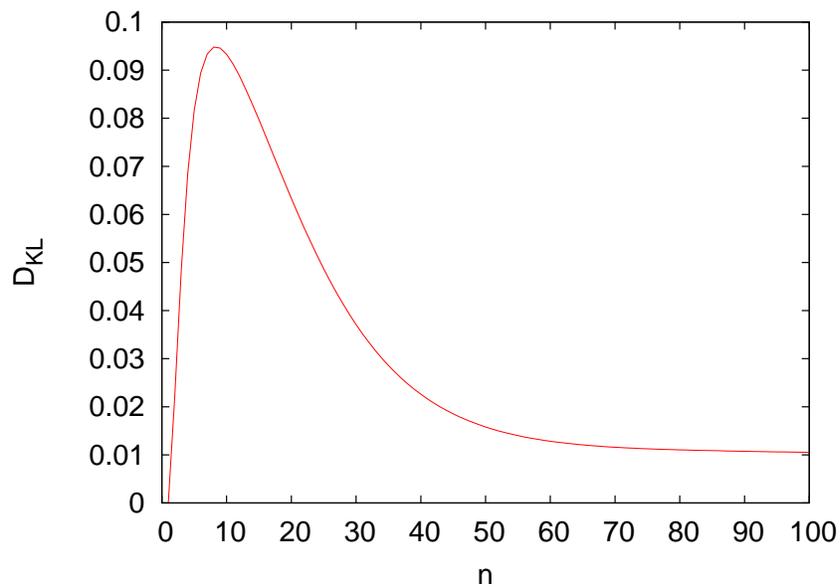}
\end{center}
\caption{\label{FigureS2} The Kullback-Leibler divergence between the
  "experimental" transition matrix $S(n\tau)$ and $S(\tau)^{n}$
  decreases after few time steps ($\tau$).}
\end{figure*}

The Figure \ref{FigureS2} reveals that the error in the transitions
computed with Chapman-Kolmogorov are far from the experimental values
during a short range of $n\tau$ close to the original lag time, but
the divergence decrease with time after few steps. In the limit $n \to
\infty$, the matrix obtained must be equal to the experimental one
(having the stationary probability distribution in each column of the
transition matrix).
\\
\\
(ii)\hspace{1cm} The {\em alanine dipeptide} is simulated with the
Langevin formalism (see details in the manuscript and in the
\emph{SI}). The continuum trajectory in the space of coordinates and
momenta is also inherently Markovian \cite{VanKampen1998Remarks2,
  Zwanzig2001Nonequilibrium2}. On the other hand, we are integrating
momenta and discretizing the coordinates space in our description,
which can result in a non-Markovian chain depending on the time step
used.  In our case the same analysis applied before provides a value
of $0.8\%$ of memory effect.

\section{Comparing with other community algorithms}

As it was introduced in the paper, one may be tempted to apply
standard algorithms for community detection in complex networks to the
problem of unveiling the topology of free energy landscapes. Several
authors have made use of current community detection tools
\cite{Radicchi2004Defining2,Donetti2004Detecting2,Girvan2002Community2,Newman2004Finding2,vanDongen20002,Newman2006From2} with diverse
degree of success. However, there is no clear conclusion about what
method is the most suitable for the problem \cite{Gfeller2007From2}. Here we
sketch the results obtained using two different approaches
\cite{vanDongen20002,Leicht2007Community2} to the community detection that have already
been used in previous related works. These results show that our
method outperforms any of the proposed algorithms both in the
computational complexity and the success in relating the CMN structure
to the FEL.

\subsection{Maximization of Modularity}

A number of community detection algorithms have been introduced in the
last years making use of the idea of modularity
\cite{Newman2006From2,Newman2004Finding2}. The idea is to find a network partition into
disconnected clusters so that a given function, the modularity of the
network partition, is maximal.  The modularity of a given partition
accounts for the probability of having edges connecting nodes
belonging to the same cluster in the network minus the expected
probability in a randomized version of the network (having the same
number of nodes and edges and preserving the strength of the
nodes). Therefore, modularity provide a way to quantify the quality of
a given network partition, {\em i.e.} the larger the modularity the
better the partitioning is.

The modularity, $Q$, for weighted and directed networks has been
introduced in ref. \cite{Leicht2007Community2,Newman2004Analysis2} and, applied to our CMN, it takes the
following form:
\begin{equation}\label{modularity}
Q=\sum_{ij}[P_{i}P_{ji}-P_{i}P_{j}] \delta (c_{i},c_{j})
\end{equation}
where the function $\delta(x,y)$ is the Kronecker delta function that takes 
value $1$ if $x=y$ and $0$ otherwise. The label $c_{i}$ accounts for 
the community where node $i$ is assigned in the network partition.

A number of optimization algorithms have been used to look for maximal
modularity partitions. We have chosen the spectral decomposition
introduced in ref.\cite{Leicht2007Community2} and a deterministic greedy search
algorithm \cite{Newman2004Fast2,Wakita2007Finding2} to compare with the results obtained
for the funnel-like CMN using the \emph{SSD} algorithm.  With these
two methods we have obtained two optimal partitions of the CMN having
$Q_{max}=0.54$ and $Q_{max}=0.53$ respectively. On the other hand, the
modularity value for the basins detected by \emph{SSD} only reaches a
value $Q=0.116$.  This would seem a bad test for the \emph{SSD},
however, as it is also shown in ref.\cite{Gfeller2007From2}, the community
structure obtained by modularity-based methods is far from the basins
shown in Figure 1C \emph{of main text}. The most approximated
decomposition is obtained by the greedy search algorithm where 11
communities were detected. In this case the most populated basin
detected for the funnel-like potential, the basin called $a$ in Figure 1C
\emph{of main text}, was split in 5 different communities.

This check was not so unwise if one observes that $Q$ is related with
the relaxation times of a stochastic Markov chain represented by the
matrix $\mathbf{S}$ in the article:

\begin{equation}\label{modularity2}
Q=\sum_{\alpha}\sum_{k \neq 1} (\tau_{k}+1)\Big (\sum_{i \in \alpha} \xi_{i}^{k} \Big )^{2},
\end{equation}

where $\alpha$ is the community index, $i$ is the node index, and
$\tau_{k}$ and $\vec{\xi}^{k}$ are the eigenvalues -relaxation times-
and eigenvectors -relaxation modes- of the stochastic matrix
$\mathbf{M} = \mathbf{S}-\mathbf{1} $.

Despite of the possible interpretation of communities based on
modularity in CMNs, the $Q$ value obtained for the \emph{SSD} basins
is quite farway from being the maximum.

\subsection{Markov Clustering algorithm}

Alternatively to the use of modularity optimization approaches, a
second type of algorithms based on random walks has been introduced to
infer the structure of complex networks. The most interesting
algorithm of this type is the Markov Clustering (MCL) approach
\cite{vanDongen20002} since it has been proved to be the best suited method
for detecting basins in CMNs \cite{Gfeller2007From2}. MCL starts with the
stochastic matrix ${\bf S}$ and iteratively alternates sequences of
two matrix operations (namely expansion and inflation) until
the convergence to an invariant matrix (from which network partition is
finally obtained) is achieved.  Expansion and inflation
transformations correspond to matrix squaring and matrix Hadamard
$p$-power respectively. These two operations are followed to a
normalization step to assure that the resulting matrix is stochastic
at the end of each iteration. The detected network partition depends
crucially on the choice of the parameter $p\geq 1$, often called
granularity. For its minimum value $p=1$ only one community is
detected, the whole network. Therefore, the success on finding the
right network community structure is achieved by a fine tuning of $p$.

\begin{figure}[h]
\begin{center}
\includegraphics[width=0.60\textwidth]{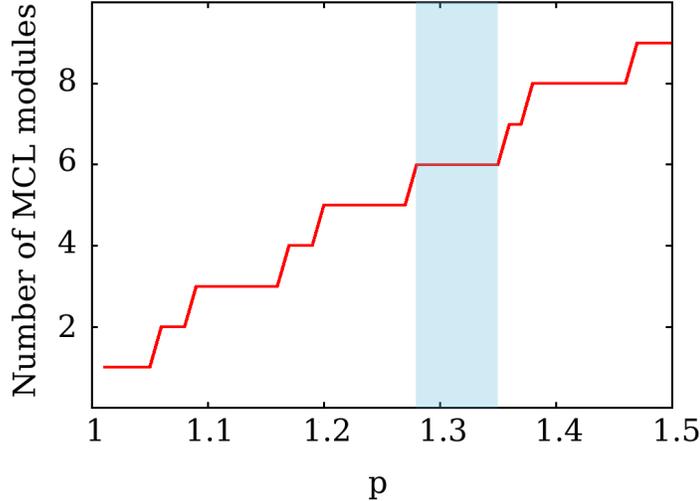}
\caption{Markov Clustering algorithm applied to the funnel-like
    potential. The figure shows the number of network clusters found
  by MCL as a function of the granularity parameter $p$. As it is
  shown there is a range of granularity values for which MCL is able
  to detect the same number of clusters as the SSD algorithm.}
\label{CML}
\end{center}
\end{figure}

We have performed the MCL algorithm over the CMN of the funnel-like
network, see Figure \ref{CML}. Unlike modularity-based algorithms, the
MCL analysis points out that the network is divided into six
communities (basins) for a small range of $p$, $1.28 \leq p \geq 1.35$
($1.5\leq p < 2.0$ is recommended in the literature), in agreement
with the partition found using the \emph{SSD} algorithm (see Table
\ref{tableS1}). However, the non-deterministic character of the method would
tempt, if we do not have prior knowledge about the structure of the
FEL, to evaluate the quality of the partitions by computing their
modularity. This would lead us to a CMN partition in disagreement with
the number of basins present in the FEL. Therefore, the free parameter
$p$ makes this algorithm unsuitable for analyzing general CMNs since
one should have enough prior knowledge about the system to distinguish
the best $p$ value.

\begin{table}[h]
  \begin{center}
  \caption{The number of nodes of SSD basins are compared with those of the MCL modules ($p=1.3$) for the funnel-like potential.}
  \multirow{9}{3mm}{\begin{sideways}SSD Basins\end{sideways}}
  \begin{tabular}{@{\vrule height 10.5pt depth4pt  width0pt}|l|cccccc|}
    \multicolumn{7}{c}{}\\
    \multicolumn{7}{c}{MCL Modules}\\
    \hline 
    & 1 & 2 & 3 & 4 & 5 & 6  \\
    \hline
    a & 56 & 3  &   &   &   &   \\
    b &    & 29 &   &   &   &   \\
    c &    &   & 54 &   &   &   \\
    d &    &   &    & 24 &  &   \\
    e &    &   &    & 1 & 49 &  \\
    f &    &   &    &   &    & 25 \\
    \hline
  \end{tabular}
    \label{tableS1}
  \end{center}
\end{table}

Finally, let us remark that both modularity optimization and MCL are
computationally far more complex than the SSD algorithm. In
particular, the modularity optimization scales as $\mathcal{O}(N^{3})$
for the spectral decomposition and as $\mathcal{O}(mN)$ when greedy
algorithm is implemented. On the other hand, MCL has a time complexity
$\mathcal{O}(N^{3})$.  To our knowledge the unique community detection
algorithm that scales linearly in time (specifically as
$\mathcal{O}(L+N)$) (with $L$ being the number of links in the
network) was proposed in ref. \cite{Wu2004Finding2}.  However, this latter method
assumes a prior knowledge of the number of network clusters and,
moreover, these cluster have to be of equal size. As a conclusion, the
comparison between SSD algorithm and standard community detection
\cite{Danon2005Comparing2} methods yields a positive assessment on the convenience
of using SSD based both on its better performance and linear time
complexity. This makes the use of the ``Stochastic steepest descent'' technique
the most appropriate method for analyzing molecular dynamics data from
systems of many degrees of freedom such as proteins.

\section{Alanine dipeptide}

\subsection{Molecular dynamics simulation}
The terminally blocked alanine peptide Ace-Ala-Nme has been modeled by
the OPLS force field \cite{Jorgensen1996Development2} with the program Gromacs 3.3.1
\cite{Gromacs2} and solvated with TIP4P waters \cite{Jorgensen1983Comparison2}.  A
Langevin MD simulation has been performed at 400K with a friction
coefficient of $5$ $ps^{-1}$. In order to set up the CMN, a single
trajectory of 250 ns has been analyzed with conformations saved every
0.01 ps (5 MD steps).

\subsection{Direct inter-basins transitions}

An extension of Table 2 of main text is shown with the whole data for
the direct transitions.

\begin{center}
 \begin{table}[h]
   \begin{center}
 \caption{Characteristic times for direct transition inter-basins.}
 \begin{tabular}{l c | l c}
 \hline
 $a\to b$ & $1/k_{ba}$ (ps) & & \\
 \hline
 $C_{7eq}\to C_{5}$ & 0.66 & $\alpha'\to C_{5}$ & 1.35 \\
 $C_{7eq}\to \alpha_{R}$ & 3.32 & $\alpha'\to C_{7eq}$ & 3.48 \\
 $C_{7eq}\to \alpha'$ & 10.80 & $\alpha'\to \alpha_{R}$ & 0.30 \\
 $C_{7eq}\to \alpha_{L}$ & 1968.34 & $\alpha'\to \alpha_{L}$ & 400.57 \\
 $C_{5}\to C_{7eq}$ & 0.33 & $C_{7ax}\to C_{5}$ & 6.63 \\
 $C_{5}\to \alpha_{R}$ & 98.25 & $C_{7ax}\to \alpha_{R}$ & 815.87 \\
 $C_{5}\to \alpha'$ & 2.11 & $C_{7ax}\to \alpha_{L}$ & 0.78 \\
 $C_{5}\to C_{7ax}$ & 393.75 & $\alpha_{L}\to C_{5}$ & 1213.40 \\
 $C_{5}\to \alpha_{L}$ & 51187.87 & $\alpha_{L}\to C_{7eq}$ & 88.24 \\
 $\alpha_{R}\to C_{5}$ & 110.28 & $\alpha_{L}\to \alpha'$ & 58.47 \\
 $\alpha_{R}\to C_{7eq}$ & 1.88 & $\alpha_{L}\to C_{7ax}$ & 4.80 \\
 $\alpha_{R}\to \alpha'$ & 0.54 &  & \\
 $\alpha_{R}\to C_{7ax}$ & 58011.74 &  & \\
 \hline
 \end{tabular}
 \end{center}
 \end{table}
\end{center}


\begin{thebibliography}{10}
\bibitem{nota1}
If the max value of $P_{j,l}$ is reached for several neighbors of $l$
  (degeneracy), we choose at random one of them.
\input{nota1}
\end{thebibliography}
\end{document}